\documentclass[prl,twocolumn,amsmath,amssymb]{revtex4}

\usepackage{graphicx}
\usepackage{bm}

\begin{document}

\title{Chandrasekhar Theory of Ellipsoidal Electromagnetic Scatterers}

\author{Peter B. Weichman}

\affiliation{BAE Systems, Advanced Information Technologies, 6 New
England Executive Park, Burlington, MA 01803}




\begin{abstract}

A number of new problems in remote sensing and identification of buried
compact metallic targets motivate the search for new models that, if
not exact, at least enable extremely rapid numerical predictions of
electromagnetic scattering/induction data. Here the elegant
Chandrasekhar theory of the electrostatics of charged ellipsoids is
used to develop an essentially exact, extremely efficient description
of low- to intermediate frequency (or late- to intermediate-time)
responses of ellipsoidal targets. Comparisons with experimental data
demonstrate that, together with a previously developed theory of the
high frequency (or early time) regime, the results serve to cover the
entire dynamic range encountered in typical measurements.

\end{abstract}


\maketitle

Exactly soluble models in the theory of EM propagation and scattering
are essentially limited to horizontally stratified or spherically
symmetric geometries, with limited results also available for
cylindrically symmetric waveguide geometries \cite{Jackson}. However,
there are a number of new problems in remote sensing and classification
of buried compact metallic targets that require a wider class of
solutions that, if not exact, at least support rapid numerical
evaluation. These include a number of longstanding economic and
humanitarian problems, such as clearance of unexploded ordnance (UXO)
from old test ranges. The most difficult technological issue is
\emph{not} the detection of such targets, but rather the ability to
distinguish between them and harmless clutter items, such as pieces of
exploded ordnance. Since clutter tends to exist at much higher density,
even modest discrimination ability leads to huge reductions in
remediation costs.

The most promising technologies, least influenced by the complexities
of the heterogeneous earth, are essentially low frequency ($< 10$ kHz)
metal detectors that emit a series of sharply terminated, well spaced
pulses, and measure the decaying induced currents during the ($\sim 25$
ms) quiet interval between pulses. At these low frequencies, detailed
spatial resolution is lost \cite{foot:gpr}, but the induced voltage
curve $V(t)$ has substantial structure, and one may hope to extract
target geometry information from it. This inference, however, is
indirect \cite{foot:pendepth} and solution of this inverse problem
requires careful comparison of predicted and measured responses. This
requires accurate models that go well beyond existing exact solutions.

A number of classes of buried objects, especially UXO, have a
cylindrical geometry. As a step in the direction of modeling such
objects, I consider here EM scattering from ellipsoids, with arbitrary
axes ${\bf a} = (a_1, a_2, a_3)$. Although fully analytic solutions are
not possible \cite{foot:vector}, it is demonstrated here that
Chandrasekhar's elegant approach to the electrostatics of
heterogeneously charged ellipsoids \cite{Chandra} enables the major
part of the computation to be performed analytically, reducing it to a
matrix diagonalization problem. The size of the matrix grows as one
pushes earlier in time, closer to pulse termination (where the high
frequency components of the pulse spectrum have not yet had a chance to
decay away), but enables an essentially exact description at
intermediate- to late-time at minimal numerical cost. As will be
demonstrated through comparisons with laboratory data on artificial
spheroidal targets, combining this approach with a complementary
early-time approach \cite{foot:earlyt} enables accurate predictions at
all time (or frequency) scales.

For clarity of exposition, the theory will be presented for nonmagnetic
target and background, $\mu = \mu_b$, where $\mu_b$ is the (uniform)
background permeability. Magnetic targets exhibit both a conductivity
and a permeability contrast, which complicates the analysis but does
not affect the basic approach. Details of this generalization will be
presented elsewhere.

For nonmagnetic systems, the Maxwell equations may be reduced, in the
frequency domain, to a single equation for the electric field
\begin{equation}
\nabla \times \nabla \times {\bf E} - \kappa^2 {\bf E} = {\bf S},
\label{1}
\end{equation}
where $\kappa^2 = \epsilon \mu k^2$, $k = \omega/c$, and source term
${\bf S} = (4\pi i \mu k/c) {\bf j}_S$ where ${\bf j}_S$ is the source
current density, representing in this case the transmitter coil. Inside
the metallic target, at all frequencies of interest here, the
dielectric function is dominated by the dc conductivity, $\epsilon =
4\pi i \sigma/\omega$. The background may be insulating or weakly
conducting, but we work in the very high contrast limit
$|\epsilon/\epsilon_b| >> 1$ \cite{foot:hcontrast}.

To obtain a closed equation, restricted to the finite target domain
$V_c$, we use the Green function approach. A background field ${\bf
E}_b$ is defined by solving (\ref{1}) for the same source ${\bf S}$,
but in the absence of a target, $\kappa^2 = \kappa_b^2$. By subtracting
this equation from (\ref{1}), and applying the background tensor Green
function ${\bf \hat G}$, defined by
\begin{equation}
\nabla \times \nabla \times {\bf \hat G}({\bf x},{\bf x}')
- \kappa_b^2 {\bf \hat G}({\bf x},{\bf x}')
= \openone \delta({\bf x}-{\bf x}'),
\label{2}
\end{equation}
where $\openone$ is the $3 \times 3$ identity matrix and the curls act
on the first index of ${\bf \hat G}$, (\ref{1}) takes the integral form
\begin{equation}
{\bf E}({\bf x}) = {\bf E}_b({\bf x}) + \int_{V_c} d^3x'
Q({\bf x}') {\bf \hat G}({\bf x},{\bf x}') \cdot {\bf E}({\bf x}'),
\label{3}
\end{equation}
where the contrast function $Q = \kappa^2 - \kappa_b^2$ vanishes
outside the target volume $V_c$. For a uniform background one obtains
${\bf \hat G} = (\openone + \kappa_b^{-2} \nabla \nabla ) g$, where
$g({\bf x},{\bf x}') = e^{i\kappa_b|{\bf x}-{\bf x}'|}/4\pi|{\bf
x}-{\bf x}'|$ is the scalar Helmholtz Green function. Restricting ${\bf
x} \in V_c$, (\ref{3}) becomes an equation for the internal field ${\bf
E}_\mathrm{int}$ alone. Given ${\bf E}_\mathrm{int}({\bf x})$, the
external field ${\bf E}_\mathrm{ext}({\bf x})$, ${\bf x} \notin V_c$,
follows by direct integration.

Over the measurement domain outside $V_c$, $\kappa_b$ can vary
strongly, encompassing soil, rock, surface plants, air, etc. The full
${\bf E}$ field is therefore unpredictable.  However, an EM induction
(EMI) measurement is sensitive only to the ``magnetic'' contribution to
${\bf E}$. Specifically, one may divide ${\bf E}$ into curl-free and
divergence free contributions, and in the high contrast limit equation
(\ref{3}) reduces to \cite{foot:waves}
\begin{equation}
{\bf E}({\bf x}) - {\bf E}_b({\bf x})
= \int_{V_c} d^3x' \frac{\kappa({\bf x}')^2 {\bf E}({\bf x}')}
{4\pi |{\bf x}-{\bf x}'|} - \nabla \varphi({\bf x}),
\label{4}
\end{equation}
in which the potential term $\nabla \varphi$, which depends on the
detailed form of $\kappa_b$, is explicitly curl free, and it follows
from (\ref{1}) that the divergence of the first term is of relative
order $\kappa_b^2/\kappa^2$ \cite{foot:poisson}. It follows as well
that (\ref{4}) enforces the boundary condition ${\bf E}_\mathrm{int}
\cdot {\bf \hat n} = O(\epsilon_b/\epsilon) \to 0$, where ${\bf \hat
n}$ is the unit surface normal. The background ${\bf E}_b$ is to be
treated as a known input here.

Remarkably, there is a straightforward procedure for solving (\ref{4})
that entirely avoids computing $\varphi$, or the noninductive part of
${\bf E}_b$. Thus, let $\{{\bf Z}_N({\bf x})\}_{N=1}^\infty$ be a
complete set of basis functions supported on $V_c$, and obeying $\nabla
\cdot {\bf Z}_N = 0$, ${\bf x} \in V_c$, and ${\bf \hat n} \cdot {\bf
Z}_N = 0$, ${\bf x} \in
\partial V_c$. One may then expand
\begin{equation}
\kappa({\bf x})^2 {\bf E}_\mathrm{int}({\bf x})
= \sum_N \zeta_N {\bf Z}_N({\bf x}).
\label{5}
\end{equation}
The key observation is that the inner product $\int_{V_c} d^3x {\bf
Z}_n \cdot \nabla f \equiv 0$ for any scalar function $f$. Therefore,
inserting (\ref{5}) into (\ref{4}) and taking the inner product on the
left with ${\bf Z}_M^*$, one obtains the matrix equation
\begin{equation}
(i\omega \hat {\cal O} - \hat {\cal L}) {\bm \zeta} = {\bm \zeta}_b,
\label{6}
\end{equation}
where the self-adjoint arrays $\hat {\cal O}$, $\hat {\cal L}$ are
defined by
\begin{eqnarray}
\hat {\cal O}_{MN} &=& -\int_{V_c} d^3x
\frac{{\bf Z}_M({\bf x})^* \cdot {\bf Z}_N({\bf x})}
{4\pi \mu \sigma({\bf x})}
\nonumber \\
\hat {\cal L}_{MN} &=& \int_{V_c} d^3x \int_{V_c} d^3x'
\frac{{\bf Z}_M({\bf x})^* \cdot {\bf Z}_N({\bf x}')}
{4\pi|{\bf x}-{\bf x}'|}
\nonumber \\
\zeta_{b,M} &=& \int_{V_c} d^3x {\bf Z}_M({\bf x})^*
\cdot {\bf E}_b({\bf x}).
\label{7}
\end{eqnarray}
Given the solution of this equation for the amplitude array ${\bm
\zeta}$, the result of an external magnetic field or EMI measurement is
also independent of $\varphi$. The curl in the magnetic field relation
${\bf B} = (i\omega)^{-1} \nabla \times {\bf E}$ annihilates $\nabla
\varphi$, while an induced voltage measurement involves a line integral
around the closed receiver loop $C_R$, which is also insensitive to any
gradient contribution.

Now, in the time domain, one is primarily interested in freely decaying
signals when ${\bf S},{\bf E}_b$, hence ${\bm \zeta}_b$, vanish. One
seeks solutions in the form of a mode expansion ${\bf E} = \sum_n A_n
{\bf E}^{(n)}({\bf x}) e^{-\lambda_n t}$, in which each (normalized)
mode shape ${\bf E}^{(n)}$ is expanded in the form (\ref{5}), and the
decay rates $\lambda_n$ and corresponding basis function coefficients
${\bm \zeta}^{(n)}$ are solutions to the generalized eigenvalue
equation,
\begin{equation}
\hat {\cal L} {\bm \zeta} = \lambda \hat {\cal O} {\bm \zeta}.
\label{8}
\end{equation}
Using mode orthogonality it can be shown that the excitation
coefficients $A_n$ are given by
\begin{equation}
A_n = I_T^{(n)} \oint_{C_T} {\bf E}^{(n)*} \cdot d{\bf l},\
I_T^{(n)} \equiv -\int_{-\infty}^0 dt e^{\lambda_n t} \partial_t I_T(t),
\label{9}
\end{equation}
in which $C_T$ is the closed transmitter loop, and $I_T(t)$ is the
transmitter pulse, terminated at $t=0$. The measured EMI voltage is
also a multiexponential sum, with amplitudes given by similar line over
$C_R$:
\begin{equation}
V_R(t) = \sum_n V_n e^{-\lambda_n t},\
V_n = A_n \oint_{C_R} {\bf E}^{(n)} \cdot d{\bf l}.
\label{10}
\end{equation}

For general targets, explicit forms for the ${\bf Z}_N$ may be hard to
come by, and for general $\sigma({\bf x})$, the integrals in (\ref{7})
must be done numerically. However, for uniform, ellipsoidal targets
both of these problems are absent. First, let ${\bf
Z}^{(\mathrm{sph})}_N$ be basis functions for the unit sphere. We will
use the forms ${\bf Z}^{(i)}_{lmp}$, $i=1,2$, defined by
\cite{foot:eharmonics}
\begin{eqnarray}
{\bf Z}^{(1)}_{lmp}({\bf x}) &=& r^{l+2p} {\bf X}_{lm}(\theta,\phi)
\nonumber \\
{\bf Z}^{(2)}_{lmp}({\bf x}) &=& \nabla \times
[(1-r^2) r^{l+2p} {\bf X}_{lm}(\theta,\phi)]
\label{11}
\end{eqnarray}
where ${\bf X}_{lm}$ are the vector spherical harmonics \cite{Jackson},
and $p = 0,1,2,\ldots$. For an ellipsoid the forms
\begin{equation}
{\bf Z}_N({\bf x}) = \sum_{\alpha=1}^3 a_\alpha
Z^{(\mathrm{sph})}_{N,\alpha}(x_1/a_1,x_2/a_2,x_3/a_3)
{\bf \hat e}_\alpha
\label{12}
\end{equation}
have the desired properties, where ${\bf \hat e}_\alpha$ are unit
vectors along the principal directions.

The key property is that ${\bf Z}_N^{(\mathrm{sph})}$ are
\emph{polynomials} (of degree $l+2p$), and hence so are ${\bf Z}_N$,
and, as advertised, the Chandrasekhar approach \cite{Chandra} allows
Coulomb integrals of the form in (\ref{4}) to be performed
analytically. The inner product of this result with another basis
function, as in (\ref{7}), is then trivial to compute. The method is
elegant, but somewhat intricate, and so will only be outlined here. The
basic result we use is \cite{Chandra}:
\begin{eqnarray}
\phi({\bf x}) &\equiv& \int d^3x'
\frac{\rho[\mu({\bf x}')]}{|{\bf x}-{\bf x}'|}
\nonumber \\
&=& \int_{\tau({\bf x})}^\infty dt
\frac{\psi(1) - \psi[\mu(t;{\bf x})]}
{\sqrt{(a_1^2+t)(a_2^2+t)(a_3^2+t)}},
\label{13}
\end{eqnarray}
where $\rho$ is any 1D function of the quantity $\mu({\bf x}) =
\sum_\alpha x_\alpha^2/a_\alpha^2 \in [0,1]$, which traces out the
family of similar ellipsoids inside $V_c$, $\psi(\mu) = \int_0^\mu
\rho(\mu') d\mu'$ is the antiderivative of $\rho$, $\mu(t,{\bf x}) =
\sum_\alpha x_\alpha^2/(a_\alpha^2 + t) \in [0,1]$, and $\tau({\bf x})
\equiv 0$ for ${\bf x} \in V_c$, and is otherwise the solution to $1 =
\sum_{\alpha=1}^3 x_\alpha^2/(a_\alpha^2 + \tau)$, ${\bf x} \notin
V_c$, which defines the family of confocal ellipsoids surrounding
$V_c$.

\begin{figure}

\includegraphics[width=0.85\columnwidth,bb = 54 145 530 667,clip]{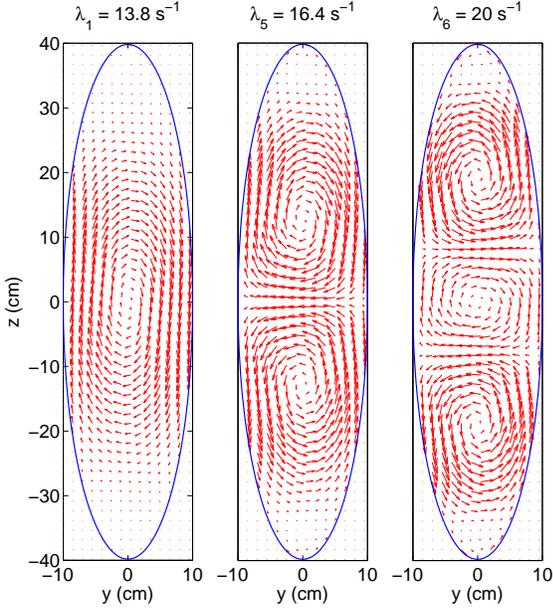}

\caption{(Color online) Three lowest order vertically circulating mode
shapes, and corresponding decay rates, for a $10 \times 10 \times 40$
cm radius aluminum prolate spheroid: $(E_y,E_z)$ are plotted in the
$x=0$ plane.}

\label{fig:vmodes}
\end{figure}

We apply this relation to compute potentials $\phi_n$ due to monomial
densities of the form $\rho_n(\mu) = (1-\mu)^n/n!$, so that
$\psi_n(1)-\psi_n(\mu) = \rho_{n+1}(\mu)$. The density in (\ref{13}) is
then a polynomial of degree $2n$. To isolate a single monomial, one
takes a combination of derivatives with respect to ${\bf x}$ and
$a_\alpha^{-2}$:
\begin{eqnarray}
D_{2{\bf l} + {\bf m}}({\bf x}) &\equiv& \int_{V_c} d^3x'
\frac{\prod_\alpha x_\alpha^{2l_\alpha + m_\alpha}}{|{\bf x}-{\bf x}'|}
\nonumber \\
&=& (-1)^{|{\bf l}+{\bf m}|} \partial_{\bf a}^{\bf l}
\left[\prod_\alpha \left(\frac{a_\alpha^2}{2}
\partial_\alpha \right)^{m_\alpha} \right]
\phi_{|{\bf l}+{\bf m}|}({\bf x})
\nonumber \\
&=& {\sum_{\bf k}}'
D_{2{\bf k}+{\bf m}}^{(2{\bf l}+{\bf m})}(\tau)
\prod_\alpha x_\alpha^{2k_\alpha + m_\alpha},
\label{14}
\end{eqnarray}
where ${\bf l} = (l_1,l_2,l_3)$ is a vector of nonnegative integers,
${\bf m} = (m_1,m_2,m_3)$ with each $m_\alpha = 0$ or 1, $|{\bf l}+{\bf
m}| = \sum_\alpha (l_\alpha + m_\alpha)$, $\partial_{\bf a}^{\bf l} =
\prod_\alpha (-1)^{l_\alpha} \partial^{l_\alpha}/\partial
(a_\alpha^{-2})^{l_\alpha}$, the primed sum is restricted to $|{\bf k}|
\leq |{\bf l}| + 1$, and
\begin{eqnarray}
D_{2{\bf k}+{\bf m}}^{(2{\bf l}+{\bf m})}(\tau)
&=& \frac{\pi a_1 a_2 a_3}{(|{\bf l}|-|{\bf k}|+1)!}
{\sum_{\bf p}}' A_{{\bf k}+{\bf m}+{\bf p}}(\tau)
\nonumber \\
&&\times\ \prod_\alpha C_{p_\alpha}^{(k_\alpha l_\alpha m_\alpha)}
a_\alpha^{2(l_\alpha+m_\alpha+p_\alpha)},
\label{15}
\end{eqnarray}
where the primed sum is over all $0 \leq p_\alpha \leq l_\alpha$, and
the combinatorial factor is
\begin{equation}
C_p^{(klm)} = \frac{(-1)^{k+l}}{k!}
\left(\begin{array}{c} l \\ p \end{array} \right)
\frac{\Gamma(\frac{1}{2}-m-p)}{\Gamma(\frac{1}{2}-m-l)}
\frac{\Gamma(k+m+p+\frac{1}{2})}{\Gamma(k+m+\frac{1}{2})}.
\label{16}
\end{equation}
Finally, the elliptic-type integrals are defined by
\begin{equation}
A_{\bf k}(\tau) = \int_\tau^\infty \frac{dt}
{\prod_\alpha (a_\alpha^2 +t)^{k_\alpha+\frac{1}{2}}}.
\label{17}
\end{equation}
These obey the iterative relation $\partial A_{\bf k}/\partial
a_\alpha^2 = -(k_\alpha + \frac{1}{2}) A_{{\bf k}+{\bf \hat
e}_\alpha}$, and evaluation may be reduced to that of $A({\bf a},\tau)
\equiv A_{\bf 0}$. With the convention $a_1 \geq a_2 \geq a_3$ one one
obtains $A({\bf a},\tau) = 2F(\theta,q)/\sqrt{a_1^2-a_3^2}$, in which
$F$ is the elliptic integral \cite{GR}, $\sin^2(\theta) =
\sqrt{(a_1^2-a_3^2)/(a_1^2 + \tau)}$ and $q^2 =
(a_1^2-a_2^2)/(a_1^2-a_3^2)$. Derivatives of $F$, and the auxiliary
elliptic integral $E(\varphi,k)$ \cite{GR}, may be expressed back in
terms of $E,F$, allowing (\ref{17}) to be evaluated by iteration. In
the case of spheroids, $a_1=a_2$ or $a_2=a_3$, fully analytic
expressions for $E,F$ are available \cite{GR}.

\begin{figure}

\includegraphics[width=0.95\columnwidth,bb = 12 131 585 640,clip]{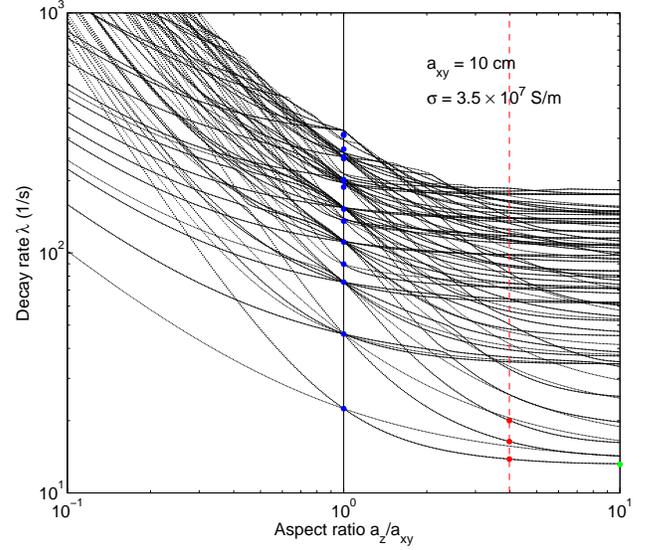}

\caption{(Color online) Decay rate spectrum vs.\ aspect ratio $\alpha =
a_z/a_{xy}$, with fixed conductivity and radius $a_{xy}$. The first 125
decay rates (computed using 232 basis functions) are plotted for each
$0.1 \leq \alpha \leq 10$. Blue dots at $\alpha=1$ show exact analytic
results for the sphere, with degeneracy effects from enhanced symmetry
evident. Red dots at $\alpha = 4$ mark the three modes shown in Fig.\
\ref{fig:vmodes} (three other nearby modes circulate around the
cylinder axis, rather than along it). Green dot at the right is the
analytic result for the lowest mode for an infinite cylinder, $\alpha
\to \infty$. More slowly increasing branches at the left, $\alpha \ll
1$, correspond to modes with current patterns circulating in the
$xy$-plane.}

\label{fig:spectrum}
\end{figure}

\begin{figure}

\includegraphics[width=0.9\columnwidth,bb=20 125 570 644,clip]{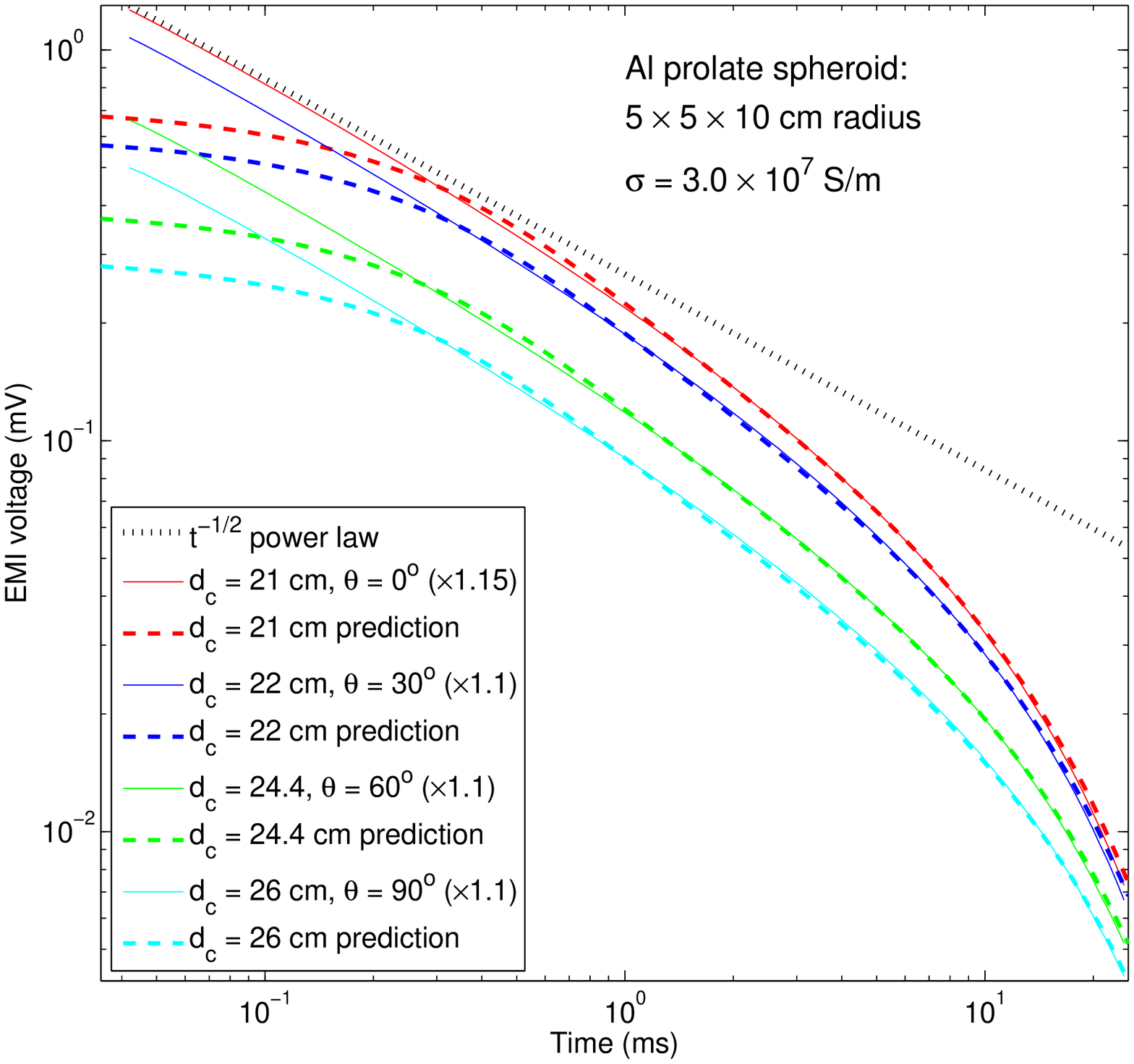}
\includegraphics[width=0.9\columnwidth,bb=20 132 570 644,clip]{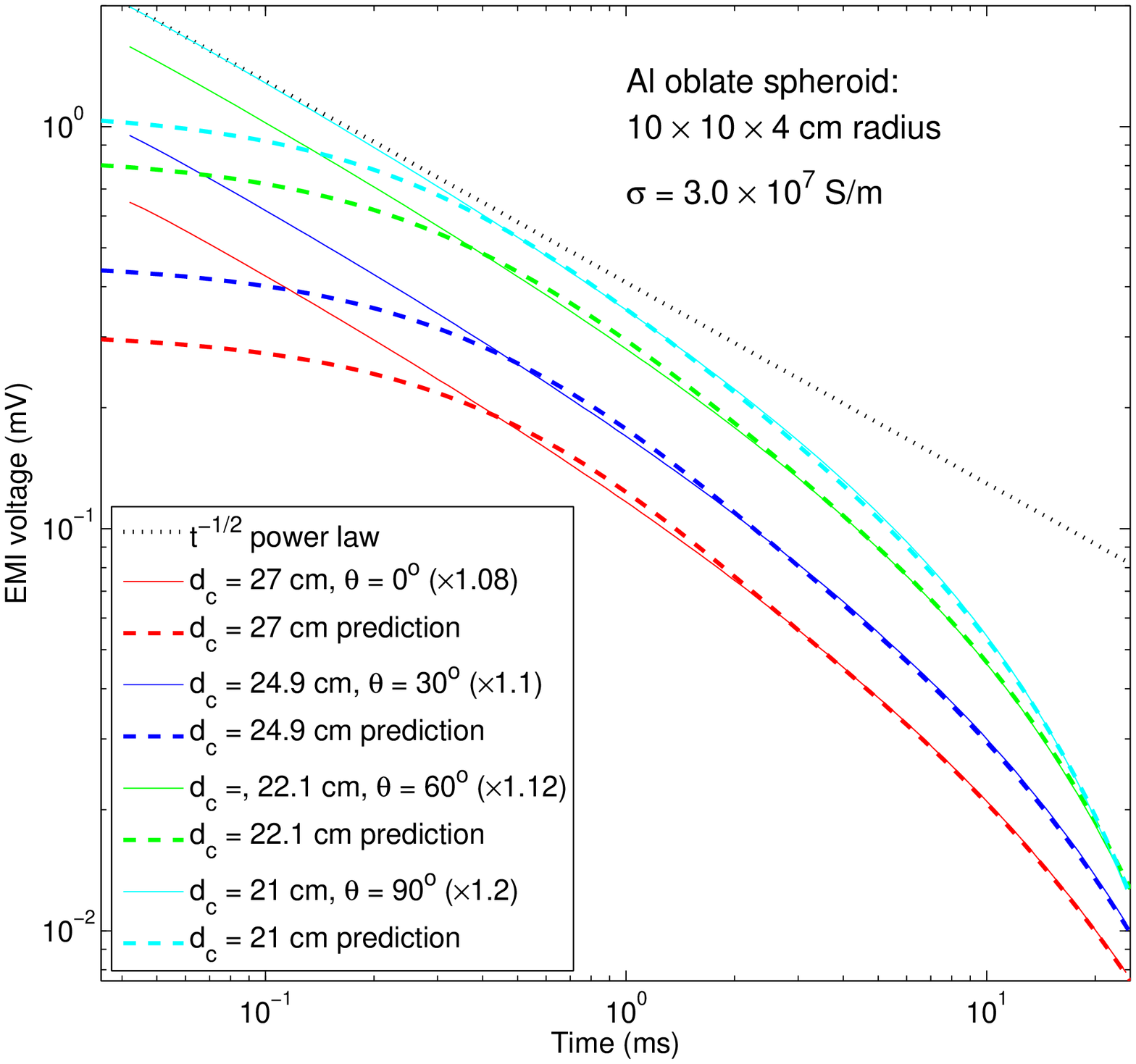}

\caption{(Color online) Comparisons of data (solid lines; taken by the
NRL TEMTADS platform \cite{TEMTADS}) and theoretical predictions
(dashed lines) for a $5 \times 5 \times 10$ cm radius aluminum prolate
spheroid (top) and a $10 \times 10 \times 4$ cm radius aluminum oblate
spheroid (bottom), at various depths-to-center $d_c$, with various
axial tilt angles $\theta$. The multipliers indicated in each legend
entry reflect a $\sim 10$\% variability in the transmitter current, and
are applied to the data to optimize the fit. Straight dashed lines show
the predicted $1/\sqrt{t}$ early time divergence \cite{foot:earlyt}.
This first principles agreement, over nearly two decades in both time
and voltage, is remarkable.}

\label{fig:datacomp}
\end{figure}

Numerical implementation proceeds as follows: (a) Use the polynomial
forms (\ref{11}), (\ref{12}) and Coulomb integrals (\ref{14}),
(\ref{15}) (with $\tau = 0$) to assemble (truncated) arrays $\hat {\cal
O}$, $\hat {\cal L}$. (b) Solve the eigenvalue equation (\ref{8}) for
the decay rates and (internal) mode shapes. (c) Use the known
transmitter geometry and pulse waveform to compute the excitation
coefficients (\ref{9}). This involves an external field computation,
the Coulomb integral of ${\bf E}_\mathrm{int}$ in (\ref{4}), which
follows from (\ref{14}), (\ref{15}) with $\tau > 0$. (d) Finally, use
the known receiver geometry to compute the voltage (\ref{10}).

Figure \ref{fig:vmodes} shows three of the computed mode shapes,
obtained using the 232 basis functions (\ref{11}) corresponding to $1
\leq l+2p \leq 7$. Figure \ref{fig:spectrum} shows the computed decay
rate spectrum for a two decade range of spheroid aspect ratios. Figure
\ref{fig:datacomp} describes comparisons with experimental data from
aluminum spheroids. The truncation loses accuracy at early time since
more rapidly decaying modes with more complex geometry are not properly
captured. However, the predictions merge smoothly with those of the
complementary early time theory \cite{foot:earlyt}, which predicts a
$1/\sqrt{t}$ divergence. The combined result yields an excellent fit
over the full dynamic range. With this level of agreement (afforded
equally by the present theory and the remarkable hardware improvement
\cite{TEMTADS}), the subtle changes in curve shape with depth and
orientation can indeed be inverted for target geometry. This will be
discussed in detail elsewhere.

The author thanks Daniel Steinhurst for providing experimental data.
This work was supported by SERDP, through the US Army Corps of
Engineers, Humphreys Engineer Center Support Activity under Contract
No.\ W912HQ-09-C-0024.


\begin{thebibliography}{}
\bibitem{foot:pendepth} Metallic EM skin depths at
    $O(10\ \mathrm{Hz})$ are $O(10\ \mathrm{cm})$. The response on
    $O(10\ \mathrm{ms})$ time scales then indirectly reflects
    target geometry on this scale. An analogy is the famous
    problem of ``hearing the shape of a drum,'' i.e., determining its
    boundary from its vibration spectrum.
\bibitem{foot:hcontrast} Typical metallic and ground conductivities are
    $\sigma \sim 10^7$ S/m, $\sigma_b \sim 0.1$ S/m, hence
    $|\epsilon/\epsilon_b| = O(10^8)$.
\bibitem{foot:poisson} An equation for $\varphi$ is obtained by
    multiplying (\ref{4}) by $\kappa_b^2$ and taking the divergence.
    However, this equation, producing finite $\varphi$ even in the
    limit $\kappa_b/\kappa \to 0$, plays no role in what follows, and
    so will not be discussed further.
\end{thebibliography}
\end{document}